\documentclass[a4paper,12pt]{article}
\usepackage{amsfonts,amsthm,amsmath,amssymb,graphicx}
\numberwithin{equation}{section}
\begin{document}

\newtheorem{definition}{Definition}[section]
\newcommand{\be}{\begin{equation}}
\newcommand{\ee}{\end{equation}}
\newcommand{\bea}{\begin{eqnarray}}
\newcommand{\eea}{\end{eqnarray}}
\newcommand{\LE}{\left[}
\newcommand{\R}{\right]}
\newcommand{\nn}{\nonumber}
\newcommand{\Tr}{\text{Tr}}
\newcommand{\N}{\mathcal{N}}
\newcommand{\G}{\Gamma}
\newcommand{\vf}{\varphi}
\newcommand{\LL}{\mathcal{L}}
\newcommand{\Op}{\mathcal{O}}
\newcommand{\HH}{\mathcal{H}}
\newcommand{\arctanh}{\text{arctanh}}
\newcommand{\up}{\uparrow}
\newcommand{\down}{\downarrow}
\newcommand{\ket}[1]{\left| #1 \right>}
\newcommand{\bra}[1]{\left< #1 \right|}
\newcommand{\ketbra}[1]{\left|#1\right>\left<#1\right|}
\newcommand{\rd}{\partial}
\newcommand{\de}{\partial}
\newcommand{\ba}{\begin{eqnarray}}
\newcommand{\ea}{\end{eqnarray}}
\newcommand{\db}{\bar{\partial}}
\newcommand{\we}{\wedge}
\newcommand{\ca}{\mathcal}
\newcommand{\lr}{\leftrightarrow}
\newcommand{\f}{\frac}
\newcommand{\s}{\sqrt}
\newcommand{\vp}{\varphi}
\newcommand{\hvp}{\hat{\varphi}}
\newcommand{\tvp}{\tilde{\varphi}}
\newcommand{\tp}{\tilde{\phi}}
\newcommand{\ti}{\tilde}
\newcommand{\ap}{\alpha}
\newcommand{\pr}{\propto}
\newcommand{\mb}{\mathbf}
\newcommand{\ddd}{\cdot\cdot\cdot}
\newcommand{\no}{\nonumber \\}
\newcommand{\la}{\langle}
\newcommand{\lb}{\rangle}
\newcommand{\ep}{\epsilon}
 \def\we{\wedge}
 \def\lr{\leftrightarrow}
 \def\f {\frac}
 \def\ti{\tilde}
 \def\ap{\alpha}
 \def\pr{\propto}
 \def\mb{\mathbf}
 \def\ddd{\cdot\cdot\cdot}
 \def\no{\nonumber \\}
 \def\la{\langle}
 \def\lb{\rangle}
 \def\ep{\epsilon}
\newcommand{\mcl}{\mathcal}
 \def\g{\gamma}
\def\tr{\text{tr}}

\begin{titlepage}
\thispagestyle{empty}

\begin{flushright}
%NORDITA-2015-137\\
EFI-16-12\\
%YITP-15-118\\
\end{flushright}
\bigskip

\begin{center}
\noindent{\large \textbf{Quantum Entanglement of Locally Excited States}\\ \large \textbf{in Maxwell Theory}}\\
\vspace{2cm}

Masahiro Nozaki $^{a}$ and Naoki Watamura $^{b}$ \\

\vspace{1cm}

{\it $^{a}$Kadanoff Center for Theoretical Physics, University of Chicago,\\
Chicago, Illinois 60637, USA \\}

{\it
$^{b}$Department of Physics
Nagoya University, \\ Nagoya 464-8602, Japan\\}

\vskip 4em
\end{center}
%%%%%%%%%%%%%%%%%%%%%%%%%%%%%
\begin{abstract}
 In 4 dimensional Maxwell gauge theory, we study the changes of (Renyi) entanglement entropy which are defined by subtracting the entropy for the ground state from the one for the locally excited states generated by acting with the gauge invariant local operators on the state. The changes for the operators which we consider in this paper reflect the electric-magnetic duality. The late-time value of changes can be interpreted in terms of electromagnetic quasi-particles. When the operator constructed of both electric and magnetic fields acts on the ground state, it shows that the operator acts on the late-time structure of quantum entanglement differently from free scalar fields.   
\end{abstract}
%%%%%%%%%%%%%%%%%%%%%%%%
\end{titlepage}

%%%%%%%%%%%%%%%%%%%%%%%%%%%%%%%
\tableofcontents
%%%%%%%%%%%%%%%%%%%%%%%%%%%%%%%
%%%%%%%%%%%%%%%%%%%%%%%%%%%%%%%%%%%%%%%%%%%%%%%%%%%%
%%%%%%%%%%%%%%%%%%%%%%%%%%%%%%%%%%%%%%%%%%%%%%%%%%%%
\section{Introduction and Summary}
%%%%%%%%%%%%%%%%%%%%%%%%%%%%%%%%%%%%%%%%%%%%%%%%%%%%
%%%%%%%%%%%%%%%%%%%%%%%%%%%%%%%%%%%%%%%%%%%%%%%%%%%%
Quantum entanglement significantly distinguishes quantum states from  classical states. It can characterize conformal field theories \cite{cft1, cft2, cft3} and topological phases \cite{tp1, tp2, tp3}. In $Gauge/Gravity$ correspondence \cite{Malda, Wit, GKP}, the structure of quantum entanglement in quantum field theories (QFTs) living on the boundary is expected to be related to the gravity in the bulk \cite{r, t1}. There are a lot of works done to reveal how the structure of quantum entanglement on the boundary corresponds to the geometry in the bulk \cite{sw1, sw2, tt2, r2, dh, t3, oog}. Therefore it is important to uncover the fundamental features which quantum entanglement possesses. (R$\acute{e}$nyi) entanglement entropy is one of the useful  quantities to investigate them. 

However the definition of (R$\acute{e}$nyi) entanglement entropy in gauge theories has subtleties \cite{kbt, yrn, D1, D2, YT, Rd1, Rd2, Cas1, Cas2, aok, Don1, Don2, tri1, tri2, Ma, Hua}.  In gauge theories, physical states have to be gauge invariant. It obeys constraints which guarantee its gauge invariance.   They make it difficult to divide the Hilbert space into subsystems $A$ and $B$ because the physical degrees of freedom in $A$ depends on the freedom in $B$ due to the constraints. Their boundary is $\partial A$.
Then the definition of (R$\acute{e}$nyi) entanglement entropy needs the precise method of dividing Hilbert space and defining the reduced density matrix $\rho_A$ which is given by tracing out the degrees of freedom in $B$,
\be
\rho_A = \tr_B \rho.
\ee 
On the other hand, the entropy in QFTs depends on a UV cutoff (ultraviolet cutoff) $\delta$ because by definition it has the UV divergence. It is given by a series expansion in conformal field theories. 
The physical degrees of  freedom around $\partial A$ have the significant effect on the terms which depend on $\delta$. The method of dividing the Hilbert space is expect to affect the degrees of freedom around $\partial A$ in the direct fashion. %Therefore after subtracting the terms which depends on $\delta$ from (R$\acute{e}$nyi) entanglement entropy, the ``renormalized" entropy is expected to avoid the subtleties which the definition of (R$\acute{e}$nyi) entanglement entropy possesses. 
In the present paper, we study the changes of (R$\acute{e}$nyi) entanglement entropy $\Delta S^{(n)}_A$ which is defined by subtracting the entropy for the ground state from the one for the locally excited state, which is defined by acting with a local operator on the ground state. Here we assume that the operator is located far from $\partial A$. We will explain it more in the next section. As in \cite{Alc, Alc2, MN1, MN2, MN3, TKO1, MN4}, their changes do not possess the UV divergence. More precisely, they measure how the local operator changes the structure of quantum entanglement. Therefore they are expected to avoid the subtleties which (R$\acute{e}$nyi) entanglement entropy has.
%The method of dividing the Hilbert space is expect to have effect the contribution of the effect at the boundary of $A$ to (R$\acute{e}$nyi) entanglement entropy.
%On the other hand, (R$\acute{e}$nyi) entanglement entropy in quantum field theories depends on the UV cutoff $\delta$. If it is expanded in $\delta$, the terms which depend on $\delta$ are affected by the effects of the subsystem boundary. If the ``renormalized" (R$\acute{e}$nyi) entanglement entropy is defined by removing  them from (R$\acute{e}$nyi) entanglement entropy, it is expected to be  independent of the method of dividing the Hilbert space. The given excited states are defined by acting with local operators on the ground state. If the excesses of (R$\acute{e}$nyi) entanglement entropy $\Delta S^{(n)}_A$ is defined by subtracting the entropy for the ground state form their entropy, $\Delta S^{(n)}_A$ does not depend on the entanglement structure of the ground state .Then $\Delta S^{(n)}_A$ can be well-defined without taking care of the method of dividing the Hilbert space. 

In this paper we study $\Delta S^{(n)}_A$ in $4d$ Maxwell gauge theory, which is a free CFT \cite{NYU}.  The previous works  \cite{MN1, MN2, MN3} show the time evolution of $\Delta S^{(n)}_A$ can be interpreted in terms of relativistic propagation of entangled quasi-particles which are created by local operators. In the free theories, the late-time value of $\Delta S^{(n)}_A$ is given by the constant, which depends on the operators.  It comes from the quantum entanglement between quasi-particles. As in \cite{MN3}, the late-time entanglement structure depends on the kind of quasi-particles. %Then it shows how they change the late-time structure of quantum entanglement. 
The authors in \cite{TKO1} show that in the specific $2d$ CFTs, it is related to the quantum dimension of the operator which acts on the ground state.
  The Authors in \cite{MN4, TH} have shown that in holographic theories the late time value of $\Delta S^{(n)}_A$ logarithmically increases similarly to the behavior of entanglement entropy for the local quenches \cite{cal, MN5}. $\Delta S^{(n)}_A$ in the finite temperature system was investigated by the authors in \cite{Cap1}. There are many works done to study the fundamental properties of $\Delta S^{(n)}_A$ \cite{recent} The time evolution and late-time value of $\Delta S^{(n)}_A$ depend on theories and the quasi-particles created by the local operator. Then we study how the structure of quantum entanglement is changed by gauge invariant local operators such as electric and magnetic fields. In particular, we study how the late-time structure of quantum entanglement depends on them. More precisely, we study how quasi-particles have the effect on the structure. %As explain later, the difference between them appears as the structure of late-time algebra for quasi-particles. 
We also study whether $\Delta S^{(n)}_A$ for gauge invariant locally excited state reflects electric-magnetic duality.  
 
\medskip  
\subsection*{Summary}
Here we briefly summarize our results in this work. We study how they change the structure of quantum entanglement by measuring the time evolution of $\Delta S^{(n)}_A$ for various gauge invariant local operators. We also study whether $\Delta S^{(n)}_A$ is invariant the electric-magnetic duality transformation.

\vspace{-3mm}
\subsubsection*{Electric-Magnetic Duality}
\vspace{-2mm}
As it will be explained later, $\Delta S^{(n)}_A$ for locally excited states are invariant under the transformation ${E}_i \rightarrow -{B}_i$ and ${B}_i \rightarrow {E}_i$ where $E$ and $B$ are electric and magnetic fields, respectively. 

\vspace{-3mm}
\subsubsection*{$E_i$ or $B_i$}
\vspace{-2mm}
If only $E_i$ or $B_i$ acts on the ground state, the time evolution of $\Delta S^{(n)}_A$ depends on the one which acts on the ground state. Because $\Delta S^{(n)}_A$ reflects the electric-magnetic duality, the entropy for $B_i$ is equal to that for $E_i$, which is the electric field along the same direction as that of magnetic field.  
$\Delta S^{(n)}_A$ for the electric and magnetic fields along the direction vertical to the entangling surface increases slower than those for fields along the directions parallel to the surface. %$\Delta S^{(n)}_A$ for the electric one along the direction vertical to the surface increases faster than that for the magnetic one along the same direction. 
However there are no difference between the effects of electromagnetic field and that of scalar one on the entanglement structure at the late time\footnote{When only a component of the electric or magnetic one acts on the ground state,  we do not consider the linear combination of them in this paper. }%Some specific linear combinations of them can affect the late-time structure of quantum entanglement. }.  

\vspace{-3mm}
\subsubsection*{Composite Operators}
\vspace{-2mm}
If the composite operator such as ${\bf B}^2$ acts on the ground state, they lead to the late-time structure of quantum entanglement in the same manner as a specific scalar operator. Then the late-time value of $\Delta S^{(n)}_A$ for that can be interpreted in terms of quasi-particles created by the scalar operator. However $\Delta S^{(n)}_A$ for some specific operators (e.g, $E_2B_3$) constructed of both electric and magnetic fields can be interpreted in terms of not the scalar quasi-particles but electromagnetic one, which is explained in section 4. Here $B_3$ ($E_3$) and $B_2$ ($E_2$) are the magnetic (electric) fields along the direction perpendicular to $\partial A$ as we will explain it later. %The late-time value can not be interpreted in terms of quasi-particles created by the scalar operator. 

\vspace{-3mm}
\subsubsection*{Late-time Algebra}
\vspace{-2mm}
 We interpret the late-time values of $\Delta S^{(n)}_A$ in terms of electromagnetic quasi-particles  created by an electromagnetic field, and derive a late-time algebra which they obey. There are commutation relations between the particles of the same kinds of fields . As we will mention later, there are also additional relations between $E_2$ ($E_3$) and $B_3$ ($B_2$), which are parallel to the entangling surface. They make the effect of electromagnetic fields different from that of scalar fields on the late-time structure of quantum entanglement.
\medskip  
\subsection*{Organization}
This paper is organized as follows. In section $2$, we will explain locally excited states and how to compute $\Delta S^{(n)}_A$ in the replica trick. We study the time evolution and late-time value of $\Delta S^{(n)}_A$ for various gauge invariant local operators in section 3. We interpret the late-time value of $\Delta S^{(n)}_A$ in terms of entangled quasi-particles  in section 4. We study how they have the effect on the late-time structure of quantum entanglement. We finish with the conclusion, future problems and the detail of propagators is included in  appendices.
%%%%%%%%%%%%%%%%%%%%%%%%%%%%%%%%%%%%%%%%%%%%%%%%%%%%
%%%%%%%%%%%%%%%%%%%%%%%%%%%%%%%%%%%%%%%%%%%%%%%%%%%%
\section{How to compute Excesses of (R$\acute{e}$nyi) Entanglement Entropy}
%%%%%%%%%%%%%%%%%%%%%%%%%%%%%%%%%%%%%%%%%%%%%%%%%%%%
%%%%%%%%%%%%%%%%%%%%%%%%%%%%%%%%%%%%%%%%%%%%%%%%%%%%
By measuring the excess of (R$\acute{e}$nyi) entanglement entropies $\Delta S^{(n)}_A$, we study how local gauge-invariant operators changes the structure of quantum entanglement in the $4d$ Maxwell gauge theory:
\be
S=-\frac{1}{4}\int d^4x F_{\mu \nu}F_{\rho \sigma}g^{\mu \rho}g^{\nu \sigma},
\ee
where $F_{\mu \nu}= \partial_{\mu} A_{\nu}- \partial_{\nu}A_{\mu}$ and $g^{\mu \nu}=diag \left(-1, 1 ,1 ,1\right)$.

In this section, we explain the definition of locally excited state and how to compute $\Delta S^{(n)}_A$ in the replica method.
%%%%%%%%%%%%%%%%%%%%%%%%%%%%%%%%%%%%%%%%%%%%%%%
\subsubsection*{The Definition of Locally Excited States}
%%%%%%%%%%%%%%%%%%%%%%%%%%%%%%%%%%%%%%%%%%%%%%%
The locally excited state is defined by acting with a gauge invariant local operator $\mathcal{O}$ such as $F^{\mu \nu}$ on the ground state:
\be \label{lestate}
\ket{\Psi} =\mathcal{N} \mathcal{O}(-t, -l, {\bf x}) \ket{0}.
\ee 
where $\mathcal{N}$ is a normalization constant and $\ket{0}$ is a gauge invariant state.
As in Figure.1, $\mathcal{O}$ is located at $t=-t,~x^1=-l$ and ${\bf x}=(x^2, x^3)$. 

% Here we explain how to compute those for excited states generated by acting local gauge invariant operators on the ground state in the replica method. 
%%%%%%%%%%%%%%%%%%%%%%%%%%%%%%%%%%%%%
\subsubsection*{Subsystem}
%%%%%%%%%%%%%%%%%%%%%%%%%%%%%%%%%%%%%
As in the previous works \cite{MN1, MN2, MN3,TKO1}, the subsystem $A$ is defined by ($t=0, x^1\ge 0$) as in Figure.1. In free theories $\Delta S^{(n)}_A$ approaches to a constant, which comes form quantum entanglement between entangled quasi-particles. In this paper we would like to study how the constant depends on gauge invariant operators. Therefore the region in Figure.1 is chosen as $A$. 
%The constant can be the quantum information which is carried by the particles.
\begin{figure}[h!]
  \centering
  \includegraphics[width=8cm]{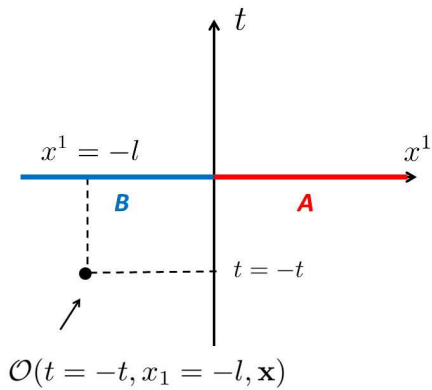}
  \caption{The location of local gauge invariant operator in Minkowski spacetime.}
\end{figure}

%We consider $4d$ Maxwell gauge theory with the Lorentzian signature ($\eta_{\mu\nu}=diag(-1, 1, 1, 1)$). 
%The locally excited states are defined by acting gauge invariant operators $\mathcal{O}$ such as $F_{\mu\nu}$  on the ground state. Here we assume that the ground state is gauge invariant state.
%The local operators $O$ is located at $t=-t$, $x^1=-l$ and ${\bf x}=(x^2, x^3)$ as in Figure. 1. 
%Then the locally excited states are defined by 

%%%%%%%%%%%%%%%%%%%%%%%%%%%%%%%%%%%%%%%%%%%%%%%%%%%%%%%%%%% 
\subsubsection*{Excesses of (R$\acute{e}$nyi) Entanglement Entropy}
%%%%%%%%%%%%%%%%%%%%%%%%%%%%%%%%%%%%%%%%%%%%%%%%%%%%%%%%%%%
Here we explain more about the definition of $\Delta S^{(n)}_A$. (R$\acute{e}$nyi) entanglement entropy for the ground state is a static quantity, which does not depend on time.  Then we define the excesses of (R$\acute{e}$nyi) entanglement entropy $\Delta S^{(n)}_A$ by subtracting $S^{(n)}_A$ for the ground state from those for locally excited states,
\be \label{exs}
\Delta S^{(n)}_A = S^{(n), EX}_A - S^{(n), G}_A, 
\ee
where $S^{(n), EX}_A$ and $S^{(n), G}_A,$ are (R$\acute{e}$nyi) entanglement entropies for the excited states in (\ref{lestate}) and the ground state $\ket{0}$, respectively. In the sense that  $\Delta S^{(n)}_A$ does not depend on $\delta$, it is a ``renormalized" (R$\acute{e}$nyi) entanglement entropy.
%%%%%%%%%%%%%%%%%%%%%%%%%%%%%%%%%%%%%%%%%%%%%
\subsection{The Replica Trick}
%%%%%%%%%%%%%%%%%%%%%%%%%%%%%%%%%%%%%%%%%%%%%
We would like to study the time evolution of $\Delta S^{(n)}_A$ in $4d$ Minkowski spacetime. However in this paper we do not directly study the changes of entanglement structure in the spacetime. Without doing so, we compute $\Delta S^{(n)}_A$ in Euclidean space by the replica trick. After that we perform the analytic continuation, which we will explain later. Then we compute the real time evolution of $\Delta S^{(n)}_A$. %Therefore we explain how to compute it in the replica method.

As in \cite{MN1, MN2, MN3, TKO1}, a reduced density matrix in Euclidean space is given by
\be \label{eulestate}
\begin{split}
\rho& %\tilde{\mathcal{N}}^2e^{-iH t}e^{- \epsilon H} \mathcal{O}(-l, {\bf x})\ket{0}\bra{0}\mathcal{O}^{\dagger}(-l, {\bf x})e^{iH t}e^{- \epsilon H}  \\
=\tilde{\mathcal{N}}^2\mathcal{O}(\tau_e, -l, {\bf x})\ket{0}\bra{0}\mathcal{O}^{\dagger}(\tau_l, -l, {\bf x}),
\end{split}
\ee
where $\tau$ is Euclidean time.
%where $\epsilon$ is introduced so that the norm of $\mathcal{O}\ket{0}$ is finite.% and we assume that $\mathcal{O}$.% is a hermitian operator\footnote{ In this paper we consider only hermitian operator. However $\mathcal{O}$ can be generalized to non-hermitian operator.}. 
By introducing a polar coordinate, $(\tau_{l, e}, -l)$ is mapped to $(r_{1, 2}, \theta_{1, 2})$ as in Figure.2.

In the replica trick, (R$\acute{e}$nyi) entanglement entropies for (\ref{eulestate}) and the ground state are respectively given by \footnote{The detail of this computation is explained in \cite{MN1, MN2, MN3, TKO1}.}
\be \label{egs}
\begin{split}
&S^{(n), EX}_A =\f{1}{1-n} \log{\left[\f{\int D\Phi \mathcal{O}^{\dagger}(r_1, \theta_1^n)\mathcal{O}(r_2, \theta^n_2)\cdots\mathcal{O}(r_1, \theta_1^1)^{\dagger}\mathcal{O}(r_2, \theta^1_2) e^{-S_n[\Phi]}}{\left(\int D\Phi \mathcal{O}(r_1, \theta_1^1)^{\dagger}\mathcal{O}(r_2, \theta^1_2) e^{-S_1[\Phi]}\right)^n}\right]}, \\
&S^{(n), G}_A =\f{1}{1-n} \log{\left[\f{\int D\Phi e^{-S_n[\Phi]}}{\left(\int D\Phi  e^{-S_1[\Phi]}\right)^n}\right]}, \\
\end{split}
\ee 
where $\theta_{1, 2}^k = \theta_{1, 2} +2(k-1)\pi $. The actions $S_n$ and $S_1$ are defined on n-sheeted geometry $\Sigma_n$ (see Figure.3) and the flat space $\Sigma_1$, respectively. %$\mathcal{O}$ are periodically located ($\theta_{1, 2}^k = \theta_{1, 2} +2(n-1)\pi $) on $\Sigma_n$ as in Figure. 3.
\begin{figure}[htb]
  \centering
  \includegraphics[width=8cm]{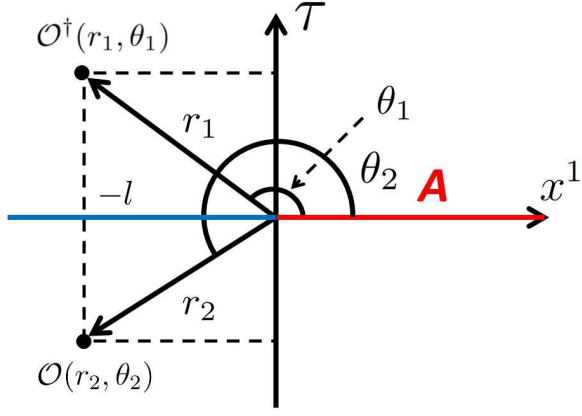}
  \caption{The location of local gauge invariant operator in Euclidean space.}
\end{figure}
By substituting (R$\acute{e}$nyi) entanglement entropies in (\ref{egs}) into (\ref{exs}), $\Delta S^{(n)}_A$ is given by
\be \label{fml}
\begin{split}
&\Delta S^{(n)}_A =\frac{1}{1-n}\log{\left[\f{\left\langle \mathcal{O}^{\dagger}(r_1, \theta^n_1)\mathcal{O}(r_2, \theta^n_2)\cdots\mathcal{O}^{\dagger}(r_1, \theta^1_1)\mathcal{O}(r_2, \theta^1_2)\right\rangle_{\Sigma_n}}{\left\langle \mathcal{O}^{\dagger}(r_1, \theta^1_1)\mathcal{O}(r_2, \theta^1_2)\right\rangle^n_{\Sigma_1}}\right]}
\end{split}
\ee

We only need to compute propagators on $\Sigma_n$ in order to compute $\Delta S^{(n)}_A$ in free field theories. The two point function of gauge fields $A_{a}$ is defined by $-\left\langle  A_{a}(r, \theta, {\bf x})A_{b}(r', \theta', {\bf x}')\right\rangle=G_{ab}\left(r, r', \theta-\theta', {\bf x}- {\bf x}' \right)$.  If we choose a specific gauge\footnote{The chosen gauge corresponds to Feynman gauge in Minkowski spacetime.}, their green functions obey the same equation of motion as that for $4d$ free massless scalar field theory, 
\be 
\begin{split} 
&\partial_r^2G^{a}_b(r, r', \theta-\theta', {\bf x}- {\bf x}')+\f{1}{r}\partial_r G^{a}_b(r, r', \theta-\theta', {\bf x}- {\bf x}') \\
&+\f{1}{r^2}\partial_{\theta}^2G^{a}_b(r, r', \theta-\theta', {\bf x}- {\bf x}')+\partial_{\bf x}^2G^{a}_b(r, r', \theta-\theta', {\bf x}- {\bf x}')=-\f{\delta^a_b\delta(r-r')\delta(\theta-\theta')\delta^2({\bf x}-{\bf x}')}{r},
\end{split}
\ee
where $a=\{\tau, x^1, x^2, x^3\}$  \footnote{$G^a_b\equiv \eta^{ac}G_{c b}$ where $\eta_{ac}= diag(1,1,1,1)$}.
 
The solution of the equation is given by
\be \label{GF}
G_{ab}\left(r, r,' \theta-\theta', {\bf x}- {\bf x}'\right)=\f{\delta_{ab}\sinh\left(\f{t_0}{n}\right)}{8n \pi^2 r r' \sinh{t_0}\left(\cosh{\left(\f{t_0}{n}\right)}-\cos{\left(\f{\theta-\theta'}{n}\right)}\right)},
\ee
where $t_0$ is defined by
\be
\cosh{t_0}=\f{r^2+r'^2+(x^2-x'^2)^2+(x^3-x'^3)^2}{2r r'}.
\ee
(\ref{GF}) has been obtained by the authors in \cite{MN1, MN2, MN3, MN4, sac, lin}.
\subsubsection*{Analytic Continuation}
After computing green functions on $\Sigma_n$ in Euclidean space, we perform the following analytic continuation,
\be \label{ana}
A_{\tau} = i A_{t},~ \partial_{\tau}= i\partial_t, ~  \tau_{l}=\epsilon-i t,~ \tau_{e}=-\epsilon -i t, 
\ee
where $\epsilon$ is a smearing parameter which is introduced to keep the norm of the excited state finite.
Analytic-continued green functions depend on $\epsilon$. We are interested in the behavior of $\Delta S^{(n)}_A$ in the limit $\epsilon \rightarrow 0$.  Their leading behavior ($\sim \mathcal{O}\left(\f{1}{\epsilon^4}\right)$) are summarized in Appendix. A.

\begin{figure}[h!]
  \centering
  \includegraphics[width=8cm]{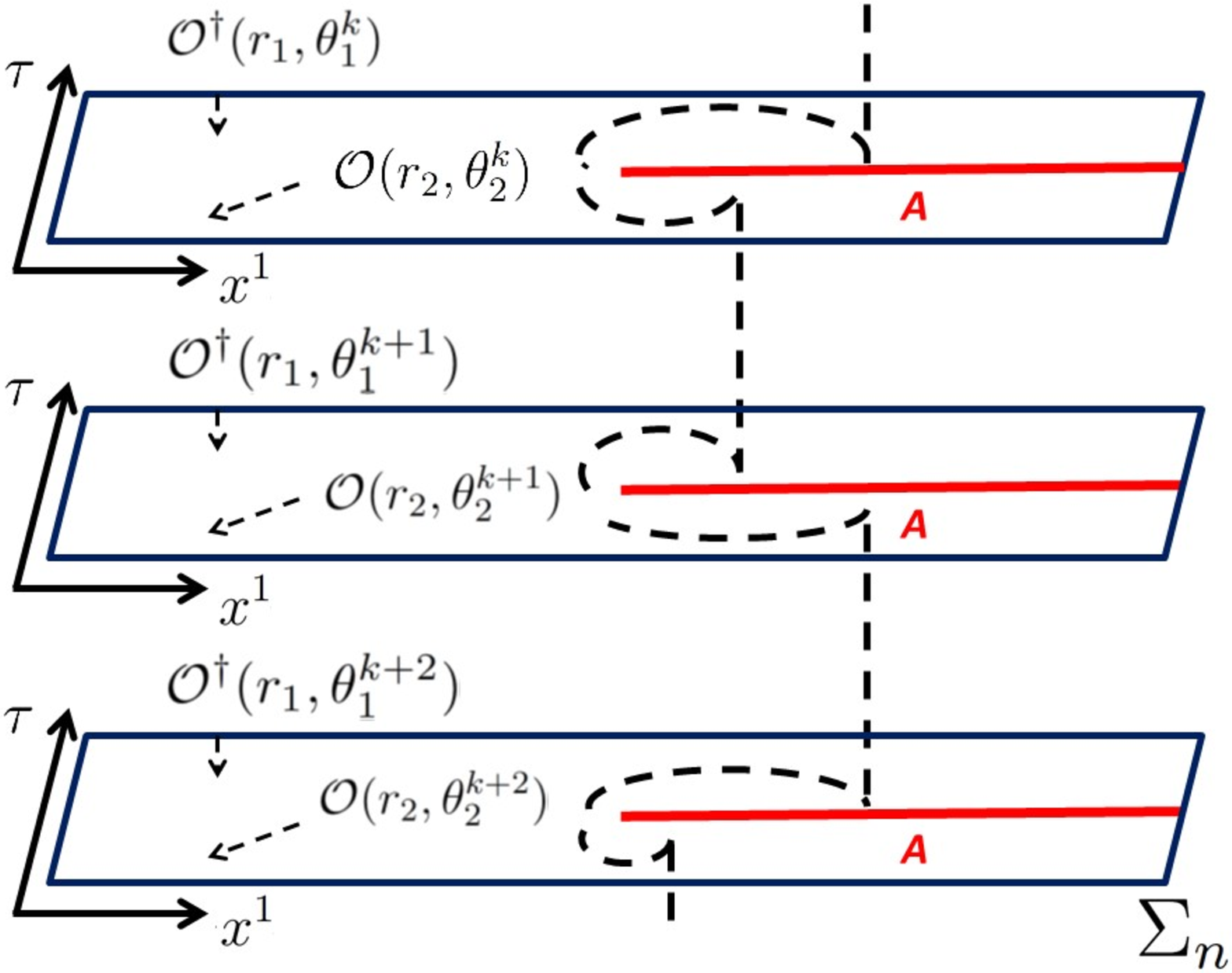}
  \caption{A picture of n-sheeted geometry $\Sigma_n$.}
\end{figure}

%%%%%%%%%%%%%%%%%%%%%%%%%%%%%%%%%%%%%%%%%%%%%
%%%%%%%%%%%%%%%%%%%%%%%%%%%%%%%%%%%%%%%%%%%%%
\section{Excesses of  (R$\acute{e}$nyi) Entanglement Entropy}
%%%%%%%%%%%%%%%%%%%%%%%%%%%%%%%%%%%%%%%%%%%%%
%%%%%%%%%%%%%%%%%%%%%%%%%%%%%%%%%%%%%%%%%%%%%
%%%%%%%%%%%%%%%%%%%%%%%%%%%%%%%%%%%%%%%%%%%%%
In this section, we study the time evolution and the late time value of $\Delta S^{(n)}_A$  in the following three cases. 
\vspace{3mm}

 (i) Only one electric or magnetic field $\mathcal{O}=E_i, B_i$ acts on the ground state. The time evolution of $\Delta S^{(n)}_A$ depends on the operator which acts on the ground state. If electromagnetic fields are changed by $F_{\mu\nu} \rightarrow \tilde{F_{\mu \nu}}=\f{1}{2}\epsilon_{\mu\nu\rho\sigma}F^{\rho \sigma}$, $\Delta S^{(n)}_A$ does not change. Here $\epsilon_{\mu\nu\rho\sigma}$ is an antisymmetric tensor. The late-time value of $\Delta S^{(n)}_A$ does not depend on the operator. It can be interpreted in terms of the quasi-particle created by a scalar operator $\phi$.

(ii) Composite operators which act on the ground state are constructed of only electric or magnetic fields such as ${\bf E}^2$ and ${\bf B}^2$. $\Delta S^{(n)}_A$ for ${\bf E}^2$ is equivalent to the entropy for ${\bf B}^2$. Then $\Delta S^{(n)}_A$ for them invariant under the electric-magnetic duality transformation. There are no differences between their effect on the (R$\acute{e}$nyi) entanglement entropy. Its late-time values can be interpreted in terms of quasi-particles, which are created by the operator constructed of massless free scalar fields $\sum_{a=1}^3(\phi^{a})^2$. Here $a$ denotes the kinds of fields. %The late-time values of $\Delta S^{(n)}_A$ for them can distinguish the effect of  ${\bf E}^2$ from that of  ${\bf B}^2$ on the late-time entanglement structure. %An effective reduced density matrix which it will be defined by (\ref{doerd}) is derived from the late time value of $\Delta S^{(n)}_A$. 

(iii) Local operators are constructed of both electric and magnetic fields such as $E_1^2+B_1^2$, $B_3E_2$ and ${\bf B}\cdot {\bf E}$. % The late-time values of $\Delta S^{(n)}_A$ for such composite operators can distinguish the effect of electric fields from the one of magnetic fields on the late-time structure.  
 The late-time value of $\Delta S^{(n)}_A$ shows that there is a significant difference between the effect of $E_1$ ($B_1$) and $E_{2, 3}$ ($B_{2, 3}$) on the late-time entanglement structure. Here $E_1$ ($B_1$) is the electric (magnetic) field along the direction vertical to the entangling surface. On the other hand, $E_{2, 3}$ ($B_{2, 3}$) is the electric (magnetic) field along the direction parallel to the entangling surface. As it will be explained in the next section, the difference comes from the commutation relation between electromagnetic quasi-particles created by $E_2$ ($E_3$) and the particles created by $B_3$ ($B_2$). %Then the late-time values of $\Delta S^{(n)}_A$ can show 
%In this section we study the excesses of (R$\acute{e}$nyi) entanglement entropy $\Delta S^{(n)}_A$ in Maxwell gauge theory in order to investigate its unique features. First locally excited states are generated by acting with a electric or magnetic field such as $E_i$ and $B_i$ on the ground state. 
%Here $i$ runs from $1$ to $3$. We study the excesses of  (R$\acute{e}$nyi) entanglement entropy which are defined by subtracting those for the ground state form those locally excited state. However their unique features is not clarified though time evolution of $\Delta S^{(n)}_A$ shows how quasi particles which are created by local operators propagates. Then we study $\Delta S^{(n)}_A$ which are defined by acting with composite operators such as $E_i E_j$ and their linear combination. In this case their features are clarified more. Particularly late-time structure of quantum entanglement which depends on features of gauge fields is clarified. In the next section, we discuss an algebra of entangled quasi particles in Maxwell gauge theory. 
\vspace{3mm}

%$E_1$ ($B_1$) is the electric (magnetic) field along the direction vertical to $\partial A$ and $E_{2, 3}$ ($B_{2, 3}$) are electric (magnetic) fields parallel to it.
%%%%%%%%%%%%%%%%%%%%%%%%%%%%%%%%%%%%%%%%%%%%%%%%%%%%%%%%%%%%%%%%%%%%%%%%%%%%%
\subsection{$\mathcal{O}= E_i $ or $B_i$}
%%%%%%%%%%%%%%%%%%%%%%%%%%%%%%%%%%%%%%%%%%%%%%%%%%%%%%%%%%%%%%%%%%%%%%%%%%%%%
Here locally excited states are defined by acting with only  $E_i$ or $B_i$ on the ground state. $\Delta S^{(n)}_A$ is given by (\ref{fml}) in the replica method with Euclidean signature. After performing the analytic continuation in (\ref{ana}) and taking the limit $\epsilon \rightarrow 0$, their time evolution is given as follows. $\Delta S^{(n)}_A$ vanishes before $t=l(>0)$, but after $t=l$, they increase. The detail of their time evolution is summarized in Table.1. After taking the late time limit $(0< l \ll t)$,  they are given by
\be
\Delta S^{(n)}_A \sim \log{2}.
\ee
Their late time value is the same as that for $\phi$ in free massless scalar field theories with any spacetime dimensions. It can be interpreted as (R$\acute{e}$nyi) entanglement entropy for maximally entangled state in $2$ qubit system. Therefore they do not show the difference between the effect of electromagnetic fields and that of free scalar one on the late-time structure of quantum entanglement. However time evolution of $\Delta S^{(n)}_A$ depends on the local operator which acts on the ground state as in Figure.4. Even at $t\sim l$, time evolution of $\Delta S^{(n)}_A$ depends on the one which acts on the ground state. 
If $\mathcal{O}=B_{2, 3}$ or $E_{2, 3}$ acts on the ground state, $\Delta S^{(n\ge 2)}_A$ is given by 
\be
\Delta S^{(n)}_A \sim \f{n}{n-1}\left(\f{3(t-l)}{4l}\right)+\cdots,
\ee
where $\cdots$ are contributions from the higher order $\mathcal{O}\left(\left(\f{t-l}{l}\right)^2\right)$.
$\Delta S^{(n\ge 2)}_A$ for $\mathcal{O}=E_1$ or $B_1$ at $t \sim l$ is given by
\be
\Delta S^{(n)}_A \sim \f{n}{n-1}\left(\f{3(t-l)^2}{4l^2}\right)+\cdots.
\ee $\cdots$ are contributions from the higher order $\mathcal{O}\left(\left(\f{t-l}{l}\right)^3\right)$.
Their time evolution shows that quasi-particles created by $E_{2, 3}$ ($B_{2, 3}$) enter the region $A$ faster than those generated by $E_1$ ($B_1$). These behaviors seem to be natural since particles created by $E_1$ ($B_1$) do not propagate along the direction parallel to $x^1$. % $\Delta S^{(n)}_A$ for $E_{2, 3}$ are different form $\Delta S^{(n)}_A$ for $B_{2, 3}$ in the intermediate region ($0<l<t$) as in Figure. 1. 

$\Delta S^{(n)}_A$ in Table.1 shows that they are invariant under the transformation,
\be \label{dtf}
F_{\mu \nu} \rightarrow \f{1}{2}\epsilon_{\mu\nu\rho\sigma}F^{\rho\sigma},
\ee
where $\epsilon_{\mu\nu\rho\sigma}$ is an anti-symmetric tensor. Under the transformation in (\ref{dtf}), the local operator $E_i$ ($B_i$) changes to $-B_i$ ($E_i$). Therefore this duality changes a locally excited state to a different one. 
\begin{figure}[h!]
  \centering
  \includegraphics[width=8cm]{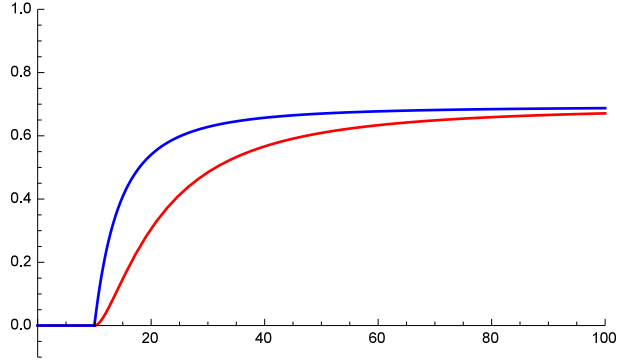}
  \caption{The time evolution of $\Delta S^{(2)}_A$ for $E_{1}$ ($B_1$) and $E_{2, 3}$ ($B_{2, 3}$). The horizontal and vertical axes correspond to time $t$ and $\Delta S^{(2)}_A$, respectively. The red and blue lines correspond to $\Delta S^{(2)}_A$  for $E_{1}$ ($B_1$) and $E_{2, 3}$ ($B_{2, 3}$), respectively.}
\end{figure}
\begin{table}[htb]
  \begin{center}
    \caption{$\Delta S^{(n)}_A$ for $\mathcal{O}$ in the region $0<l\le t$}
    \begin{tabular}{|l|c|} \hline
      $\mathcal{O}$ & $\Delta S^{(n)}_A$ \\ \hline  \hline 
      $E_1$ or $B_1$ & $\frac{1}{1-n}\log \left(\left(-\frac{(l+t)^2 (l-2 t)}{4 t^3}\right)^n+\left(\frac{(t-l)^2 (l+2 t)}{4 t^3}\right)^n\right)$  \\ \hline
      $E_{2, 3}$ &$ \frac{1}{1-n}\log \left(\left(\frac{-l^3-3 l t^2+4 t^3}{8 t^3}\right)^n+\left(\frac{l^3+3 l t^2+4 t^3}{8 t^3}\right)^n\right) $\\ \hline
 %     $B_1$ & $\frac{1}{1-n}\log \left(\frac{\left(\frac{1}{2} (t-l)^2\right)^n+\left(\frac{1}{2} (l+t)^2\right)^n}{\left(l^2+t^2\right)^n}\right)$  \\ \hline
    %  $B_{2, 3}$ &$ \frac{1}{1-n}\text{log}\left(\frac{\left(\frac{t^3-l^3}{64  t^3 }+\frac{(t-l)^2}{128  t^2 }\right)^n+\left(\frac{l^3+t^3}{64  t^3 }+\frac{(l+t)^2}{128  t^2 }\right)^n}{\left(\frac{l^2+t^2}{64  t^2 }+\frac{1}{32  }\right)^n}\right)$  \\ \hline
    \end{tabular}
  \end{center}
\end{table}
%%%%%%%%%%%%%%%%%%%%%%%%%%%%%%%%%%%%%%%%%%%%%%%%%%%%%%%%%%%%%%%%%%%%%%%%%%%%%%
\subsection{Composite Operators Constructed of Only Electric or Magnetic Fields}
%%%%%%%%%%%%%%%%%%%%%%%%%%%%%%%%%%%%%%%%%%%%%%%%%%%%%%%%%%%%%%%%%%%%%%%%%%%%%%
The excited states which we consider here are generated by acting with the following operators: (a) $E_i E_j$ or $B_i B_j$, (b) ${\bf E}^2$ or ${\bf B}^2$. 
We study the time evolution and the late-time value of $\Delta S^{(n)}_A$ for them.
%%%%%%%%%%%%%%%%%%%%%%%%%%%%%%%%%%%%%%%%%%%%%%%%%%%%%%%%%%%%%%%%%%%%%%%%%%%%%%%%
\subsubsection{$\mathcal{O} = E_i E_j$ or $B_i B_j$}
%%%%%%%%%%%%%%%%%%%%%%%%%%%%%%%%%%%%%%%%%%%%%%%%%%%%%%%%%%%%%%%%%%%%%%%%%%%%%%%%
First we consider $\Delta S^{(n)}_A$ for the excited states generated by acting with $E_iE_j$ or $B_iB_j$ on the ground state.
When $i=j$, the late time-value of $\Delta S^{(n)}_A$ is given by 
\be \label{nbl}
\Delta S^{(n)}_A =- \f{1}{1-n}\log{\f{4^n}{2^n+2}}.
\ee   
It is the same as that of $\Delta S^{(n)}_A$ for $\phi^2$ in the massless free scalar field theories as in \cite{MN1, MN2}. 
Therefore the late-time value of $\Delta S^{(n)}_A$ ((R$\acute{e}$nyi) entanglement  entropy of operator) can be interpreted in terms of entangled quasi-particles created by $\phi^2$. 

When $i\neq j$, $\Delta S^{(n)}_A$ at the late time is given by 
\be\label{neqlate}
\Delta S^{(n)}_A =\log{4},
\ee
which can be interpreted as maximum (R$\acute{e}$nyi) entanglement entropy for $\rho_A =\f{1}{4}diag (1, 1, 1, 1)$. It is the same as $\Delta S^{(n)}_A$ for the excited state given by acting  with the operator $\phi^a \phi^b$ on the ground state. Here $a, b$ denote the kind of scalar fields, and $a \neq b$. They are two kinds of massless free scalar fields.
The time evolution of $\Delta S^{(n)}_A$ for them is summarized in Table.2. 
Table.2 shows $\Delta S^{(n)}_A$ is invariant under the transformation in (\ref{dtf}).
%Late-time values of $\Delta S^{(n)}_A$ for $E_i E_j$ or $B_i B_j$ can not show the difference between the effects of electric and magnetic fields on the late-time entanglement structure.

%In the next subsection we study $\Delta S^{(n)}_A$ for the states generated by acting on the ground state with linear combinations of the products of electric fields and magnetic field in order to investigate      their features. In particular, we intend to study how the structure of quantum entanglement at the late time depends on electric and magnetic fields. 

\begin{table}[htb]
  \begin{center}
    \caption{$\Delta S^{(n)}_A=\f{1}{1-n}\log{\left[\f{N_1+N_2+N_3}{D_1}\right]}$ for $\mathcal{O}$ in the region $0<l<t$}\scalebox{0.7}{  
    \begin{tabular}{|l|l|l|l|l|} \hline
      $\mathcal{O}$ & $D_1$ & $N_1$ & $N_2$ & $N_3$  \\ \hline  
$E^2_1$ or  $B^2_1$ & $\left(2 \left(\frac{1}{16}\right)^2\right)^n$ & $\left(2\left( f_1\right)^2\right)^n$ & $\left(2\left( f_2\right)^2\right)^n$ & $2^{2 n} \left(f_1f_2\right)^n$  \\ \hline
$E^2_{2, 3}$ or $B^2_{2, 3}$&$\left(2 \left(\frac{1}{16}\right)^2\right)^n$& $\left(2\left( f_3\right)^2\right)^n$ & $\left(2\left( f_4\right)^2\right)^n$ & $2^{2 n} \left(f_3f_4\right)^n$   \\ \hline
%$$ & $$ & $f_1=\frac{(l-5 t) (l+t)}{128 t^2}$ & $f_2=\frac{(l-t) (l+5 t)}{128 t^2}$ & $$ \\ \hline
%$B^2_1$ & $\left(2 \left(\frac{l^2+t^2}{32 t^2}\right)^2\right)^n$ & $\left(2 \left(\frac{(l+t)^2}{64 t^2}\right)^2\right)^n$ & $\left(2 \left(\frac{(t-l)^2}{64 t^2}\right)^2\right)^n$ & $2^{2 n} \left(\frac{(l+t)^2}{64 t^2}\frac{(t-l)^2}{64 t^2}\right)^n$ \\ \hline
%$B^2_{2,3}$ & $\left(2 \left(\frac{\left(\frac{l^2}{t^2}+3\right)}{64} \right)^2\right)^n$ & $\left(2 (f_3)^2\right)^n$ & $\left(2 (f_4)^2\right)^n$ & $2^{2 n} \left(f_3\right)^n \left(f_4\right)^n$ \\ \hline
 %& $$ & $f_3=\frac{2 \left(l^3+t^3\right)+t (l+t)^2}{128 t^3}$ & $f_4=-\frac{(l-t) \left(2 l^2+l t+3 t^2\right)}{128 t^3}$ & $$ \\ \hline
$E_1 E_{2,3}$ or $E_1 B_{2,3}$ & $$ & $$ & $$ & $$ \\ 
or $B_1 E_{2,3}$ or $B_1 B_{2,3}$ & $\left(\frac{1}{16}\right)^{2 n}$ & $\left(f_1\right)^n \left(f_3\right)^n$ & $\left(f_2\right)^n \left(f_4\right)^n$ & $\left(f_2\right)^n \left(f_3\right)^n+\left(f_1\right)^n \left(f_4\right)^n$  \\ \hline
$E_2 E_3$ or $B_2B_3$ & $\left(\frac{1}{16}\right)^{2 n}$ & $\left(f_3\right){}^{2 n}$ & $\left(f_4\right){}^{2 n}$ & $2 \left(f_3\right){}^n \left(f_4\right){}^n$ \\ \hline
$E_{2}B_{2}$ or $E_{3}B_{3}$ & $\left(\left(\frac{1}{16}\right)^2\right)^n$ & $\left(f_3\right){}^{2 n}$ & $\left(f_4\right){}^{2 n}$ & $2 \left(f_3\right){}^n \left(f_4\right){}^n$ \\ \hline
%$B_1B_{2, 3}$ o& $\left(\frac{\left(\frac{l^2}{t^2}+3\right)}{64} \right)^n \left(\frac{l^2+t^2}{32 t^2}\right)^n$ & $f_3^n \left(\frac{(l+t)^2}{64 t^2}\right)^n$ & $f_4^n \left(\frac{(t-l)^2}{64 t^2}\right)^n$ & $f_3^n \left(\frac{(t-l)^2}{64 t^2}\right)^n+f_4^n \left(\frac{(l+t)^2}{64 t^2}\right)^n$ \\ \hline
%$B_2B_3$ & $\left(\frac{\left(\frac{l^2}{t^2}+3\right)}{64} \right)^{2 n}$ & $f_3^{2 n}$ & $f_4^{2 n}$ & $2 f_3^n f_4^n$ \\ \hline
$E_1B_1$ & $\left(\frac{1}{16}\right)^{2 n}$ & $\left(f_1\right){}^{2 n}$ & $\left(f_2\right){}^{2 n}$ & $2 \left(f_1\right){}^n \left(f_2\right){}^n$ \\\hline
%$E_1B_{2, 3}$ & $\left(-\frac{\frac{l^2}{t^2}+3}{16\cdot 64}\right)^n$ & $\left(\frac{f_3 \left((l-2 t) (l+t)^2\right)}{64 t^3}\right){}^n$ & $\left(-\frac{f_4 \left((l+2 t) (t-l)^2\right)}{64 t^3}\right){}^n$&$\left(-\frac{f_3 \left((l+2 t) (t-l)^2\right)}{64 t^3}\right){}^n+\left(\frac{f_4 \left((l-2 t) (l+t)^2\right)}{64 t^3}\right){}^n$ \\\hline
%$E_{2, 3}B_{2, 3}$ & $\left(\frac{\left(\frac{l^2}{t^2}-5\right) \left(\frac{l^2}{t^2}+3\right)}{64\ 64}\right)^n$ & $\left(f_1 f_3\right){}^n$&$\left(f_2 f_4\right){}^n$ &$\left(f_2 f_3\right){}^n+\left(f_1 f_4\right){}^n$\\\hline\hline
%Functions & $f_1$ & $f_2$ & $f_3$ & $f_4$  \\ \hline
Functions & $f_1=-\frac{(l-2 t) (l+t)^2}{64 t^3}$ & $f_2=\frac{(l+2 t) (l-t)^2}{64 t^3}$ & $f_3=\frac{l^3+3 l t^2+4 t^3}{128 t^3}$ & $f_4=\frac{-l^3-3 l t^2+4 t^3}{128 t^3}$  \\ \hline
    \end{tabular}}
  \end{center}
\end{table}
%%%%%%%%%%%%%%%%%%%%%
\subsubsection{$\mathcal{O}= {\bf E}^2$ or ${\bf B}^2$}
%%%%%%%%%%%%%%%%%%%%%
In order to study whether $E_1$ acts on the late-time structure of quantum entanglement differently from $E_{2, 3}$\footnote{The effect of $E_1$ can be different from that of $E_{2, 3}$ on the structure since we choose $t=0, x^1\ge 0$ as the subsystem $A$.}
, we study the late-time value of $\Delta S^{(n)}_A$ for the given locally excited state:
\be\label{EE}
\ket{\Psi}= \mathcal{N} {\bf E}^2 (-t, -l, {\bf x})\ket{0}.
\ee
Before studying its late-time value, we comment on its time evolution.
Before $t=l$, $\Delta S^{(n)}_A$ for the state in (\ref{EE}) vanishes and after $t=l$, it increases. Its time evolution is summarized in Table.3. 

After $t=l$, as in Table.3, $\Delta S^{(n)}_A$ is given by 
\be\label{FEE}
\begin{split}
\Delta S^{(n)}_A = 
\f{1}{1-n}\log{\left[\f{N_1+N_2+P_1+P_2+P_3}{D}\right]},
\end{split}
\ee
where $D$, $N_i$ and $P_i$ are defined in Table.3. 
If we take the late time limit ($0<l\ll t$),  the ratios of $P_i$ and $N_i$ to $D$ reduce to constant numbers \cite{MNF},
\be \label{NPR}
\begin{split}
&\left(\f{N_1}{D}\right)^{\f{1}{n}}=\left(\f{N_2}{D}\right)^{\f{1}{n}}= 4^{-1}, ~~\left(\f{P_1}{D}\right)^{\f{1}{n}} =\left(\f{P_2}{D}\right)^{\f{1}{n}}=\left(\f{P_3}{D}\right)^{\f{1}{n}}=\left(\frac{1}{6}\right), \\
\end{split}
\ee 
where we ignore the higher order contribution $\mathcal{O}\left(\f{l}{t}\right)$. Amazingly, The sum of them is $1$. Therefore if the effective reduced density matrix is defined by
\be \label{doerd}
\Delta S^{(n)}_A =\f{1}{1-n}\log{\left[\tr_A \left(\rho^{e}_A\right)^n\right]},
\ee
then the matrix is given by
\be \label{rdmee}
\begin{split} 
\rho^e_A = \f{1}{24}diag(6, 4, 4, 4, 6).
\end{split}
\ee
The excess of $n-$th (R$\acute{e}$nyi)  entanglement entropy, entanglement entropy and Min entropy are respectively given by
\be \label{entropyee}
\begin{split}
&\Delta S^{(n)}_A = \frac{1}{n-1}\log {\left(\frac{12^n}{3\cdot 2^n+2\cdot 3^n}\right)}, \\
&\Delta S_A = \frac{\log (24)}{2}, \\
&\Delta S^{(\infty)}_A=\log {4},
\end{split}
\ee 
which can  be interpreted in terms of quasi-particles created by $\left(\phi^1\right)^2+\left(\phi^2\right)^2+\left(\phi^3\right)^2$, which is constructed of three kinds of free scalar fields. Therefore, there are no differences between the effect of $E_1$ and that of $E_{2, 3}$ on the late-time structure of quantum entanglement. As in the Table.2, $\Delta S^{(n)}_A$ for ${\bf E}^2$ is equivalent to that for ${\bf B}^2$. Therefore they is the electric-magnetic duality invariant. %not be interpreted in terms of quasi-particles created by free scalar fields. Then they can distinguish the effect of $E_1$ from that of $E_{2, 3}$ on the late-time entanglement structure. The difference between them appears as the one between their commutation relation as it will be explained in the next section.
%%%%%%%%%%%%%%%%%%%%%%%%%%%%%%%%%%%
%\subsubsection{$\mathcal{O}= {\bf B}^2$}
%%%%%%%%%%%%%%%%%%%%%%%%%%%%%%%%%%
%Here, to study how different the effect of $B_1$ is from that of $B_{2, 3}$, let's study the time evolution and the late-time value of $\Delta S^{(n)}_A$ for the excited state defined by
%\be
%\ket{\Psi}= \mathcal{N} {\bf B}^2(-t, -l, {\bf x})\ket{0}.
%\ee 
%Before $t=l$, $\Delta S^{(n)}_A$ for the state vanishes and after $t=l$, its time evolution is summarized in Table. 3. 

%After $t=l$, it is given by (\ref{FEE}).
%If we take the late time limit, $\left(\f{N_i}{D}\right)^{\f{1}{n}}$ and $\left(\f{P_i}{D}\right)^{\f{1}{n}} $ reduce to constants similarly to (\ref{NPR}). Then the effective reduced density matrix $\rho^{e}_A$ is given by
%\be \label{erdmbb}
%\begin{split}
%\rho^e_A = \f{1}{44}\begin{pmatrix}
%11&0&0&0&0 \\
%0&9&0&0&0 \\
%0&0&4&0&0 \\
%0&0&0&9&0 \\
%0&0&0&0&11 \\
%\end{pmatrix}
.
%\end{split}
%\ee
%Entropies for it  are respectively given by
%\be \label{entropybb}
%\begin{split}
%&\Delta S^{(n)}_A = \frac{1}{n-1}\log {\left(\frac{44^n}{4^n+2\cdot 9^n+2\cdot 11^n}\right)}, \\
%&\Delta S_A = \log \left(2\cdot \left(\frac{2}{3}\right)^{9/11} \sqrt{11}\right), \\
%&\Delta S^{(\infty)}_A=\log {4},
%\end{split}%
%\ee 
%which show the difference between the effects of magnetic fields on the late-time structure. 
\begin{table}[htb] 
  \begin{center}
    \caption{$\Delta S^{(n)}_A=\f{1}{1-n}\log{\left[\f{N_1+N_2+P_1+P_2+P_3}{D_1}\right]}$ for $\mathcal{O}$ in the region $0<l<t$}
    \scalebox{0.6}{\begin{tabular}{|l|l|l|l|l|l|l|} \hline
      $\mathcal{O}$ & $D_1$ & $N_1$ & $N_2$    &$P_1$&$P_2$&$P_3$ \\ \hline
     ${\bf B}^2 or {\bf E}^2$ & $\left(2\cdot 3 \left(\frac{1}{16}\right)^2\right)^n$ & $\left(2f_1^2+2\cdot 2f_3^2\right)^n$&$\left(2f_2^2+2\cdot 2f_4^2\right)^n$&$2^{2n}f_1^n f_2^n$ &$2^{2n}f_3^n f_4^n$&$2^{2n}f_3^n f_4^n$ \\ \hline
 $B_1^2+E_1^2$ & $\left(2\cdot 2 \left(\frac{1}{16}\right)^2\right)^n$ & $\left(2\cdot 2 f_1^2\right)^n$&$\left(2\cdot 2 f_2^2\right)^n$&$2^{2n}f_1^n f_2^n$ &$2^{2n}f_1^n f_2^n$&$0$ \\ \hline
$E_2 B_3$& $\left(2^2 \left(\frac{1}{4\cdot 8}\right)^2\right)^n$ & $\left(2 \left(g_1\right)^2+2\left(g_3\right)^2\right)^n$ & $\left(2 \left(g_2\right)^2+2\left(g_4\right)^2\right)^n$&$2^{2 n} \left(g_2\right)^n \left(g_3\right)^n$&$2^{2 n} \left(g_1\right)^n \left(g_4\right)^n$&$0$ \\
or $E_3 B_2$ &&&&&& \\ \hline
$F^{\mu\nu}F_{\mu\nu}$   & $\left(2\cdot 2 \left(\frac{1}{16}\right)^2+2\cdot 4^2 \left(\frac{1}{4\cdot 8}\right)^2\right)^n$ & $\left(2\cdot2f^2_1+2\cdot4^2g_1\cdot g_3\right)^n$ &  $\left(2\cdot2f^2_2+2\cdot4^2g_2\cdot g_4\right)^n$&$2\cdot2^{2 n} \left(f_1\right)^n \left(f_2\right)^n$&$2\cdot4^{2 n} \left(g_1\right)^n \left(g_2\right)^n$& $2\cdot4^{2 n} \left(g_3\right)^n \left(g_4\right)^n$ \\ 
or ${\bf B}\cdot{\bf E}$ &&&&&& \\ \hline
$B_2E_3-B_3E_2$   & $\left(4\cdot 2 \left(\frac{1}{4\cdot 8}\right)^2\right)^n$ & $\left(2\cdot2g^2_3+2\cdot2g^2_1\right)^n$ &  $\left(2\cdot2g^2_4+2\cdot2g^2_2\right)^n$&$2^{2 n} \left(g_3\right)^n \left(g_2\right)^n$&$2^{2 n} \left(g_1\right)^n \left(g_4\right)^n$&$0$ \\ \hline
Functions & $g_1=\frac{(l+t)^3}{4\cdot 64 t^3}$ & $g_2=\frac{(t-l)^3}{4\cdot 64 t^3}$&$g_3=\frac{(l+t) \left(l^2-4 l t+7 t^2\right)}{4\cdot 64 t^3}$&$g_4=\frac{(t-l) \left(l^2+4 l t+7 t^2\right)}{4\cdot 64 t^3}$&& \\ \hline
    \end{tabular}}
  \end{center}
\end{table}
%%%%%%%%%%%%%%%%%%%%%%%%%%%%%%%%%%%%%%%
%%%%%%%%%%%%%%%%%%%%%%%%%%%%%%%%%%%%%%%
\subsection{Composite Operators Constructed of Both Electric and Magnetic Fields}
%%%%%%%%%%%%%%%%%%%%%%%%%%%%%%%%%%%%%%%
%%%%%%%%%%%%%%%%%%%%%%%%%%%%%%%%%%%%%%%
In the previous two subsection we study how the entanglement structure changes at the late time if either electric or magnetic fields act on the ground state. However we do not uncover how it changes at the late time when both of them act on the ground state. Here we study $\Delta S^{(n)}_A$ for (a) $E_1^2+B_1^2$, (b) $E_i B_j$ and (c) $F^{\mu\nu}F_{\mu \nu}$ and ${\bf B}\cdot {\bf E}$, which can show that $E_i$ and $B_i$ act on the late-time structure of quantum entanglement differently from scalar fields such as $\phi^a$.  % We do not uncover the difference between the effect of electric fields and that of magnetic fields on the quantum entanglement at the late time.
%Then we study them here. 
%%%%%%%%%%%%%%%%%%%%%%%%%%
\subsubsection{$E^2_1+B^2_1$}
%%%%%%%%%%%%%%%%%%%%%%%%%%
Here in order to study whether there are differences between the effects of electric and magnetic fields on the late-time structure of quantum entanglement, we study $\Delta S^{(n)}_A$ for the following excited state:
\be
\ket{\Psi}= \mathcal{N} \left(E^2_1+B^2_1\right)(-t, -l, {\bf x})\ket{0}.
\ee

Before investing the late time value of $\Delta S^{(n)}_A$, let's study the time evolution of $\Delta S^{(n)}_A$.
$\Delta S^{(n)}_A$ vanishes before $t=l$. After $t=l$, its time evolution is summarized in Table.3.
If you take the late time limit $t \rightarrow \infty$, the late time value of $\Delta S^{(n)}_A$ reduces to the (R$\acute{e}$nyi) entanglement entropy whose effective reduced density matrix is given by
\be \label{rdmeb2}
\begin{split}
\rho^e_A =\f{1}{4}diag\left(1, 1, 1, 1\right)
.
\end{split}
\ee
Its entropies are given by
\be
\begin{split}
&\Delta S^{(n)}_A =\Delta S_A = \Delta S^{(\infty)}_A= \log{4}.\\
%&\log{4}, \\
%&\Delta S^{(\infty)}_A=\log {\left(\frac{5}{2}\right)}.
\end{split}
\ee
It shows there are no differences between the effects of electric and magnetic fields on the late-time structure. 
%%%%%%%%%%%%%%%%%%%%%%%%%%%%%%%%%%%%%%%
\subsubsection{$E_iB_j$}
%%%%%%%%%%%%%%%%%%%%%%%%%%%%%%%%%%%%%%%
Here let's find out how the operators constructed of both electric and magnetic fields, $E_i B_j$, affect the late-time structure of quantum entanglement.
The late-time values of $\Delta S^{(n)}_A$ for $E_i B_j$ except for $E_2 B_3$ and $E_3 B_2$ are the same as (\ref{neqlate}). Their time evolution is summarized in Table.2. 

On the other hand, after $t=l$ the time evolution of $\Delta S^{(n)}_A$ for $E_2 B_3$ or  $E_3 B_2$
is summarized in Table.3. We can see that it has the electric-magnetic duality from the Table.3. The late-time value of $\Delta S^{(n)}_A$ is given by (R$\acute{e}$nyi) entanglement entropy whose reduced density matrix is given by %\footnote{We checked $\Delta S^{(n)}_A$ up to $n=4$.}
\be \label{erdmbe}
\begin{split}
\rho^e_A= \f{1}{64}diag(25, 7, 7, 25).
\end{split} 
\ee

Its entropies are given by
\be
\begin{split}
&\Delta S^{(n)}_A =-\frac{\log \left(2^{1-6 n} \left(7^n+25^n\right)\right)}{n-1}, \\
&\Delta S_A =\log \left(\frac{64}{5\cdot 5^{9/16} 7^{7/32}}\right), \\
&\Delta S^{(\infty)}_A=2\log{\left(\f{8}{5}\right)}. \\
\end{split}
\ee 

It shows how different the effect of $E_1$ ($B_1$) is from that of $E_{2, 3}$ ($B_{2, 3}$) on the structure. The value can not be interpreted in terms of quasi-particles created by scalar fields such as $\phi^a \phi^b$. 
As we will explain later, in the entangled quasi-particle interpretation, there is a commutation relation between the quasi-particle created by $E_2$ ($B_2$) and that by $B_3$ ($E_3$).

\subsubsection{${\bf B}\cdot {\bf E}$ and $F_{\mu \nu}F^{\mu \nu}$ and $B_2E_3-B_3E_2$}
We finally study $\Delta S^{(n)}_A$ for more complicated operators, ${\bf B}\cdot {\bf E}$, $F_{\mu \nu}F^{\mu \nu}$ and $B_2E_3-B_3E_2$. Before $t=l$, $\Delta S^{(n)}_A$ for them vanish, but after $t=l$, they increases. The detail of them is summarized in Table.3\footnote{$\Delta S^{(n)}_A$ is commuted by the green functions in Appendix.B.}. It shows that $\Delta S^{(n)}_A$ for ${\bf B} \cdot {\bf E}$ is the same as that for $F^{\mu\nu}F_{\mu\nu}$.
 %Precisely we study how these operators change the late-time structure of quantum entanglement. 
%If $\mathcal {O}={\bf B }\cdot{\bf E}$ and $F^{\mu\nu}F_{\mu\nu}$ act on the ground state, the excesses of (R$\acute{e}$nyi) entanglement entropy are given by
%\be \label{rees}
%\begin{split}
%&\Delta S^{(2), {\bf B}\cdot {\bf E}}_A=\log \left(\frac{1444}{337}\right), ~~
%\Delta S^{(2), F^{\mu \nu}F_{\mu\nu}}_A=\log \left(\frac{3872}{861}\right),\\
%&\Delta S^{(3), {\bf B}\cdot {\bf E}}_A=\frac{1}{2} \log \left(\frac{6859}{445}\right)), ~~
%\Delta S^{(3), F^{\mu \nu}F_{\mu\nu}}_A=\log \left(176 \sqrt{\frac{11}{19331}}\right),\\
%&\Delta S^{(4), {\bf B}\cdot {\bf E}}_A=\log \left(76 \sqrt[3]{\frac{19}{160729}}\right)~~
%\Delta S^{(4), F^{\mu \nu}F_{\mu\nu}}_A=\frac{1}{3} \log \left(\frac{2725888}{42129}\right).\\
%\end{split}
%\ee
%where $\Delta S^{(n), {\bf B}\cdot{\bf E}}_A$ and $\Delta S^{(n), F^{\mu\nu}F_{\mu\nu}}$ are the excesses of (R$\acute{e}$nyi) entanglement entropy for ${\bf B}\cdot{\bf E}$ and $ F^{\mu\nu}F_{\mu\nu}$, respectively.

The effective reduced density matrices for ${\bf B} \cdot {\bf E}$ ($F^{\mu \nu }F_{\mu \nu}$ ) and $B_2E_3-B_3E_2$ are given by
\be\label{erdmco}
\begin{split}
&\rho^e_A =\f{1}{192}diag\left(30, 30, 16, 16, 49, 49, 1, 1\right) , ~~~~\text{ for}~~ \mathcal{O}={\bf B} \cdot {\bf E}~~(F^{\mu\nu}F_{\mu\nu}),\\
&\rho^e_A =\f{1}{128}diag\left(50, 50, 7, 7, 7, 7\right) , ~~~~~~~~~~~~~~~~\text{ for}~~ \mathcal{O}=B_2E_3-B_3E_2.\\
\end{split}
\ee

Entropies  are respectively given by
\be
\begin{split}
&\Delta S^{(n) }_A=\frac{1}{n-1}\log {\left(\frac{2^{6 n-1} 3^n}{16^n+30^n+49^n+1}\right)}, \\
&\Delta S_A=\log \left(\frac{32 \cdot  3^{11/16}\sqrt[48]{\frac{2}{7}}}{7\cdot 5^{5/16}}\right), \\
&\Delta S^{(\infty)}_A =\log {\left(\f{192}{49}\right)}.\\
\end{split}
\ee
which are for $\mathcal{O}={\bf B} \cdot {\bf E}$ (or $F^{\mu\nu}F_{\mu\nu}$), and
\be
\begin{split}
&\Delta S^{(n)}_A=-\frac{\log \left(2^{1-7 n} \left(2\cdot 7^n+50^n\right)\right)}{n-1}, \\
&\Delta S_A=\log \left(\frac{64 \left(\frac{2}{7}\right)^{7/32}}{5\cdot 5^{9/16}}\right), \\
&\Delta S^{(\infty)}_A =2\log{ \left(\frac{8}{5}\right)}, \\
\end{split}
\ee
which are for $\mathcal{O}=B_2E_3-B_3E_2$. As we will explain in the next section, they can be reproduced by using a late-time algebra which electromagnetic quasi-particles obey. %It is derived from the results in sections 3.1, 3.2, 3.3.1 and 3.3.2.
%%%%%%%%%%%%%%%%%%%%%%%%%%%%%%
\section{A Late-time Algebra}
%%%%%%%%%%%%%%%%%%%%%%%%%%%%%%
We interpret the late-time value of $\Delta S^{(n)}_A$ in terms of quasi-particles. More precisely, let's interpret the effective reduced density matrix in (\ref{doerd}) in terms of quasi-particles. The effective reduced density matrix for the excited state generated by a composite operator $\mathcal{O}(-t, -l, {\bf x})$ is defined by 
\be \label{rerdm}
\Delta S^{(n)}_A =\f{1}{1-n}\log{\left[\tr_A\left(\rho^e_A\right)^n\right]}= \f{1}{1-n}\log{\left[\tr_A\left(\hat{\mathcal{N}}^2 \mathcal{O}\ket{0}\bra{0}\mathcal{O}^{\dagger}\right)^n\right]},
\ee
where $\hat{\mathcal{N}}$ is a normalization constant. 
The operator $\mathcal{O}$ is assumed to be constructed of electric and magnetic fields\footnote{Here $\rho^e_A$ is not the same as the reduced density matrix for the locally excited state. It is for a ``effective" state $\hat{\mathcal{N}}\mathcal{O}\ket{0}$. It is different form the ``original" locally excited state.}. 
As in \cite{MN1, MN2, MN3, MN4}, these fields can be decomposed into left moving and right moving electromagnetic quasi-particles as follows,
\be \label{qps}
\begin{split}
&E_{i}=E_{i}^{L \dagger}+E_{i}^{R \dagger}+ E_{i}^{L}+E_{i}^R, \\
&B_{i}=B_{i}^{L \dagger}+ B_{i}^{R \dagger}+B_{i}^{L}+B_{i}^R, \\
\end{split}
\ee
where since we take $x^1\ge 0$ as $A$ in this paper, left-moving and right-moving quasi-particles correspond to particles included in $B$ and $A$ at late time, respectively. 
The ground state for them is defined by
\be 
\begin{split}
&E_{i}^{L, R} \ket{0}_{L, R}=B_{i}^{L, R}\ket{0}_{L, R}=0, \\
&\ket{0}=\ket{0}_L \otimes \ket{0}_R. \\
\end{split}
\ee

The late-time algebra which quasi-particles obey is given by
\be \label{lqp}
\begin{split}
\left[E_{i}^{L,R }, E_{j}^{L, R \dagger}\right]= C\delta_{i j}, \\
\left[B_{i}^{L,R}, B_{j}^{L, R \dagger}\right]= C\delta_{i j}, \\
\end{split}
\ee
which is obtained so that the results by the replica trick are reproduced. Here $C$ is a real number\footnote{The redefinition of quasi-particles can absorb the constant $C$. }. 
 In the gauge theory in addition to (\ref{qps}), we need the following commutation relation for different particles:
\be\label{lqp2}
\begin{split}
\left[E_{3}^{L,R }, B_{2}^{L, R \dagger}\right]= X_{R, L}, ~\left[E_{2}^{L,R}, B_{3}^{L, R \dagger}\right]= Y_{R, L},
\end{split}
\ee
where $X_{R, L}$ and $Y_{R, L}$ are given by
\be \label{lqp3}
\begin{split}
&X_R=-X_L=Y_L=-Y_R, \\
&X_{R, L}^2=Y_{R, L}^2 = \f{9}{16}C^2.
\end{split}
\ee
Here $X_{R, L}$ and $Y_{R, L}$ are real numbers\footnote{We find the correspondence between propagators and commutation relations. The commutations can be defined by the late time limit of propagators. We will discuss the detail of the correspondence in \cite{MNF}. When we use this correspondence, $X_L= -\f{3}{4}C$. }. %The commutation relation in (\ref{lqp2}) makes the effects of electromagnetic fields different from that of scalar fields on the late-time structure.
 The commutation relation between electric (magnetic) quasi-particles is  determined so that the effective density matrices computed by (\ref{lqp}) are consistent with (\ref{rdmee}) respectively. The relation for the quasi-particles by $E_1$ should be the same as that for $B_1$ so that the effective density matrix in (\ref{rerdm}) reproduces $\Delta S^{(n)}_A$ for the matrix in (\ref{rdmeb2}). That between quasi-particles generated by $E_2$ ($E_3$) and those by $B_3$ ($B_2$) reproduces the matrix in (\ref{erdmbe}).
 We also check that $\Delta S^{(n)}_A$ for $\mathcal{O}={\bf B}\cdot{\bf E}$ $(F^{\mu\nu}F_{\mu \nu}$), $B_3E_2-B_2E_3$ are reproduced by using the commutation relation in (\ref{lqp}) and (\ref{lqp2}). 

%These commutation relation describes the probability that quasi-particles are included in $A$ and $B$ at the late time.  The relation for left and right moving quasi-particles corresponds to the probability that they are included in $B$ and  that in $A$, respectively. As explained in section 3.1, the particles created by $E_1$ and $B_1$ can not propagate along the direction perpendicular to the entangling surface. Then it seems to be rational that the probability that the quasi-particle created by $E_1$ ($B_1$) is included in $A$ or $B$  is smaller than that for $E_{2, 3}$ and $B_{2, 3}$  even at the late time.  

%The commutation in (\ref{lqp}) shows that the probability that the electric quasi-particle for $E_1$ is included in $A$ is four times greater than the probability that the magnetic one for $B_1$ is included in $A$. We can check it by comparing the second component of $\rho^e_A$  for $F^{\mu\nu}F_{\mu \nu}$ in (\ref{erdmco}) with its fourth component. 

The relation in (\ref{lqp2}) shows that the effect of fields along the direction vertical to $\partial A$ is significantly different from that along the direction parallel to $\partial A$ on the late-time structure. It makes the effects of electromagnetic fields different from that of free scalar fields on the late-time structure of quantum entanglement. 

\section{Conclusion and Future Problems}
We also studied how gauge invariant operators such as $E_i$, $B_i$ and the composite operators constructed of them changes the structure of quantum entanglement by studying $\Delta S^{(n)}_A$. We studied whether $\Delta S^{(n)}_A$ for locally excited states created by gauge invariant local operators reflects the electric magnetic duality. $\Delta S^{(n)}_A$, which we studied in this paper, is invariant under the duality transformation. If only $E_i$ or $B_i$ acts on the ground state, without taking the late time limit, the time evolution of $\Delta S^{(n)}_A$ depends on them. Due to the duality, $\Delta S^{(n)}_A$ for $E_i$ is equal to that for $B_i$. Around $t = l$, $\Delta S^{(n)}_A$ for $E_{2, 3}$ ($B_{2, 3}$) increases slower than that for $E_1$ ($B_1$). However they can not show the difference between the effects of electromagnetic fields and that of scalar fields on the late-time structure because the late-time values of $\Delta S^{(n)}_A$ for them can be interpreted in terms of quasi-particle created by scalar fields. 

On the other hand, the late-time values of $\Delta S^{(n)}_A$ for the specific operators constructed of   both electric and magnetic fields can not be interpreted in terms of quasi-particles by scalar fields. They show that there are differences between the effects of electromagnetic and that of scalar fields on the late-time structure of quantum entanglement. If their late-time values are interpreted in terms of electromagnetic quasi-particles in (\ref{qps}), there are commutation relations between $E_2$ ($E_3$) and $B_3$ ($B_2$), which make the effect of electromagnetic field significantly different from that of scalar fields on the  late-time structure. The effect of $E_1$ and $B_1$ on the late-time structure is different from that of $E_{2, 3}$ and $B_{2, 3}$.
%However late-time values of $\Delta S^{(n)}_A$ can not show the difference between their effects. On the other hand, $\Delta S^{(n)}_A$ for composite operators can detect it even if the late time limit is taken. We interpreted the late time value of $\Delta S^{(n)}_A$ in terms of quasi-particles which obey the late-time algebra in (\ref{lqp}) and (\ref{lqp2}). The algebra shows the difference between the effect of the fields along the direction perpendicular to $\partial A$ and that along the direction parallel to $\partial A$. As in (\ref{erdmco}), the probability that the electric quasi-particle generated by $E_1$ is included in $A$ is greater than the one that the magnetic one created by $B_1$ is included in $A$. The effect of $E_1$ and $B_1$ on the late-time structure is significantly different from that of $E_{2, 3}$ and $B_{2, 3}$ because there are commutation relation between $E_2$ ($B_2$) and $B_3$  ($E_3$).
%The algebra shows that $E_1$ and $B_1$ stronger affect on late-time entanglement structure than $E_{i \neq 1}$ and $B_{i \neq 1}$ respectively. It shows that quasi-particles created by $E_1$ or $B_1$ more frequently propagate to $A$ than that by $E_{i \neq 1}$ and $B_{i \neq 1}$ respectively. $E_i$ have the stronger effect on the late-time structure than that of $B_i$. There are non-trivial quantum entanglement between $B_2$ and $E_3$ ($E_2$ and $B_3$).  

\subsubsection*{Future Problem}
We finish with comments on a few of future problems:

\begin{itemize}
\item In this paper we only consider $4d$ Maxwell gauge theory which has conformal symmetry. $D (\neq 4)$ dimensional Maxwell gauge theory is not a CFT. Therefore it is interesting to generalize the analysis in $4d$ Maxwell theory to that in the theories with general dimensions. 

\item We expect that the structure of the late-time algebra depends on the spacetime  dimension $D$. Then it is also interesting to study it in general dimensions.

%\item In this paper we propose a late-time algebra which quasi-particles obey. However it is not obvious that  the effect of electric fields on the late-time structure of quantum entanglement is different from that of magnetic fields. Then it is one of the future problems to reveal why the difference between them appears.
\end{itemize}

%%%%%%%%%%%%%%%%%%%%%%%%%%%%%%%
\subsection*{Acknowledgments}
%%%%%%%%%%%%%%%%%%%%%%%%%%%%%%%
MN thanks Tadashi Takayanagi for useful discussions and comments on this paper.  MN and NW thank Pawel Caputa, Tokiro Numasawa, Shunji Matsuura and  Akinori Tanaka for useful comments on this work. 
%%%%%%%%%%%%%%%%%%%%%%%%%%%%%%%%%%%%%%%%%%%%%%%%%%%%%%%%%%%%%%%%%%%%%%%%%%%
\appendix
%%%%%%%%%%%%%%%%%%%%%%%%%%%%%%%%%%%%%%%%%%%%%%%%%%%%%%%%%%%%%%%%%%%%%%%%%%%
\section{Green Functions}
The relation between $E_i, B_i$ and field strengths which are defined in Euclidean space is given by
\be 
E_i=-i F_{\tau i}, ~B_1=-F_{23}, ~B_2=F_{13}, ~B_3=-F_{12}.
\ee 
The analytic continued green functions are defined by
\be
\begin{split}
&\left\langle E_1(\theta)E_1(\theta')\right\rangle=F_{E1E1}(\theta-\theta'), \\
& \left\langle E_2(\theta)E_2(\theta')\right\rangle=\left\langle E_3(\theta)E_3(\theta')\right\rangle=F_{E2E2}(\theta-\theta'), \\
&\left\langle B_1(\theta)B_1(\theta')\right\rangle=F_{B1B1}(\theta-\theta'), \\
& \left\langle B_2(\theta)B_2(\theta')\right\rangle=\left\langle B_3(\theta)B_3(\theta')\right\rangle=F_{B2B2}(\theta-\theta'), \\
& \left\langle E_2(\theta)B_3(\theta')\right\rangle=F_{E2B3}(\theta-\theta'), \\
& \left\langle B_3(\theta)E_2(\theta')\right\rangle=F_{B3E2}(\theta-\theta'), \\
& \left\langle E_3(\theta)B_2(\theta')\right\rangle=F_{E3B2}(\theta-\theta'), \\
& \left\langle B_2(\theta)E_3(\theta')\right\rangle=F_{B2E3}(\theta-\theta'), \\
\end{split}
\ee
If the limit $\epsilon \rightarrow 0$ is taken, their leading terms for $n=1$ are given by
\be \label{dpge}
\begin{split}
&F_{E1E1}(\theta_1-\theta_2) \sim \frac{1}{16 \pi ^2 \epsilon ^4}, \\
&F_{E2E2}(\theta_1-\theta_2)\sim \frac{1}{16 \pi ^2 \epsilon ^4}, \\
&F_{B1B1}(\theta_1-\theta_2)\sim \frac{1}{16 \pi ^2 \epsilon ^4}, \\
&F_{B2B2}(\theta_1-\theta_2)\sim \frac{1}{16 \pi ^2 \epsilon ^4}.
\end{split}
\ee
That for the arbitrary $n$ in $0<t<l$ at the  are given by (\ref{dpge})

That for the arbitrary $n$ in $0<l \le t$ at the  are given by
\be \label{dpgnl}
\begin{split}
&F_{E1E1}(\theta_1-\theta_2)=F_{E1E1}(\theta_2-\theta_1) \sim-\frac{(l-2 t) (l+t)^2}{64 \pi ^2 t^3 \epsilon ^4}, \\
&F_{E2E2}(\theta_1-\theta_2)=F_{E2E2}(\theta_2-\theta_1)\sim \frac{l^3+3 l t^2+4 t^3}{128 \pi ^2 t^3 \epsilon ^4}, \\
&F_{B1B1}(\theta_1-\theta_2)=F_{B1B1}(\theta_2-\theta_1)\sim -\frac{(l-2 t) (l+t)^2}{64 \pi ^2 t^3 \epsilon ^4}, \\
&F_{B2B2}(\theta_1-\theta_2)=F_{B2B2}(\theta_2-\theta_1)\sim \frac{l^3+3 l t^2+4 t^3}{128 \pi ^2 t^3 \epsilon ^4},\\
&F_{E2B3}(\theta_1-\theta_2)=F_{E2B3}(\theta_2-\theta_1) \sim \frac{3 (t-l) (l+t)}{128 \pi ^2 t^2 \epsilon ^4},\\
&F_{B3E2}(\theta_1-\theta_2)=F_{B3E2}(\theta_2-\theta_1) \sim \frac{3 (t-l) (l+t)}{128 \pi ^2 t^2 \epsilon ^4},\\
&F_{E3B2}(\theta_1-\theta_2)=F_{E3B2}(\theta_2-\theta_1) \sim \frac{3 (l-t) (l+t)}{128 \pi ^2 t^2 \epsilon ^4},\\
&F_{B2E3}(\theta_1-\theta_2)=F_{B2E3}(\theta_2-\theta_1) \sim \frac{3 (l-t) (l+t)}{128 \pi ^2 t^2 \epsilon ^4},\\
%%%%%%%%%%%%%%%%%%%%%%%%%%%%%%%%%%%%%%%%%%%%%%%%%%%%%%%%%%%%%%%%%%%%%%%%%%%%
&F_{E1E1}(\theta_1-\theta_2+2\pi)=F_{E1E1}(\theta_2-\theta_1-2\pi) \\
&=F_{E1E1}(\theta_1-\theta_2-2(n-1)\pi)=F_{E1E1}(\theta_2-\theta_1+2(n-1)\pi)\sim \frac{(l-t)^2 (l+2 t)}{64 \pi ^2 t^3 \epsilon ^4}, \\
&F_{E2E2}(\theta_1-\theta_2+2\pi)=F_{E2E2}(\theta_2-\theta_1-2\pi) \\
&=F_{E2E2}(\theta_1-\theta_2-2(n-1)\pi)=F_{E2E2}(\theta_2-\theta_1+2(n-1)\pi)\sim-\frac{l^3+3 l t^2-4 t^3}{128 \pi ^2 t^3 \epsilon ^4}, \\
&F_{B1B1}(\theta_1-\theta_2+2\pi)=F_{B1B1}(\theta_2-\theta_1-2\pi) \\
&=F_{B1B1}(\theta_1-\theta_2-2(n-1)\pi)=F_{B1B1}(\theta_2-\theta_1+2(n-1)\pi)\sim \frac{(l-t)^2 (l+2 t)}{64 \pi ^2 t^3 \epsilon ^4},\\
&F_{B2B2}(\theta_1-\theta_2+2\pi)=F^{(n,l)}_{B2B2}(\theta_2-\theta_1-2\pi)\\
&=F_{B2B2}(\theta_1-\theta_2-2(n-1)\pi)=F^{(n,l)}_{B2B2}(\theta_2-\theta_1+2(n-1)\pi)\sim-\frac{l^3+3 l t^2-4 t^3}{128 \pi ^2 t^3 \epsilon ^4}, \\
&F_{E2B3}(\theta_1-\theta_2+2\pi)=F_{E2B3}(\theta_2-\theta_1-2\pi) \\
&=F_{E2B3}(\theta_1-\theta_2-2(n-1)\pi)=F_{E2B3}(\theta_2-\theta_1+2(n-1)\pi)\sim\frac{3 (l-t) (l+t)}{128 \pi ^2 t^2 \epsilon ^4} \\ 
&F_{B3E2}(\theta_1-\theta_2+2\pi)=F_{B3E2}(\theta_2-\theta_1-2\pi)\\
&=F_{B3E2}(\theta_1-\theta_2-2(n-1)\pi)=F_{B3E2}(\theta_2-\theta_1+2(n-1)\pi) \sim \frac{3 (l-t) (l+t)}{128 \pi ^2 t^2 \epsilon ^4} ,\\
&F_{E3B2}(\theta_1-\theta_2+2\pi)=F_{E3B2}(\theta_2-\theta_1-2\pi) \\
&=F_{E3B2}(\theta_1-\theta_2-2(n-1)\pi)=F_{E3B2}(\theta_2-\theta_1+2(n-1)\pi) \sim \frac{3 (t-l) (l+t)}{128 \pi ^2 t^2 \epsilon ^4}, \\
&F_{B2E3}(\theta_1-\theta_2+2\pi)=F_{B2E3}(\theta_2-\theta_1-2\pi)\\
&=F_{B2E3}(\theta_1-\theta_2-2(n-1)\pi)=F_{B2E3}(\theta_2-\theta_1+2(n-1)\pi)\sim\frac{3 (t-l) (l+t)}{128 \pi ^2 t^2 \epsilon ^4}.\\
\end{split}
\ee
The contribution of the other propagators is much smaller than them in (\ref{dpgnl}).
\section{Other Bases}
We introduce new bases $\mathcal{O}_{1,2}=\f{E_2\pm B_3}{2}$, $\mathcal{O}_{3,4}=\f{E_{3}\pm B_{2}}{2}$.
Their green functions in ($0<t<l$) are given by
\be
\begin{split}
&\left\langle \mathcal{O}_1 \mathcal{O}_1 \right\rangle (\theta_1-\theta_2) \sim 
 \frac{1}{4\cdot8 \pi ^2 \epsilon ^4}, \\
&\left\langle \mathcal{O}_2 \mathcal{O}_2 \right\rangle (\theta_1-\theta_2) \sim 
 \frac{1}{4\cdot8 \pi ^2 \epsilon ^4}, \\
&\left\langle \mathcal{O}_3 \mathcal{O}_3 \right\rangle (\theta_1-\theta_2) \sim 
 \frac{1}{4\cdot8 \pi ^2 \epsilon ^4}, \\
&\left\langle \mathcal{O}_4 \mathcal{O}_4 \right\rangle (\theta_1-\theta_2) \sim 
 \frac{1}{4\cdot8 \pi ^2 \epsilon ^4}, \\
\end{split}
\ee
and those in ($0<l\le t$) are given by
\be
\begin{split}
&\left\langle \mathcal{O}_1 \mathcal{O}_1 \right\rangle (\theta_1-\theta_2)= \left\langle \mathcal{O}_1 \mathcal{O}_1 \right\rangle (\theta_2-\theta_1)\sim 
 \frac{(l+t) \left(l^2-4 l t+7 t^2\right)}{4\cdot64 \pi ^2 t^3 \epsilon ^4}, \\
&\left\langle \mathcal{O}_2 \mathcal{O}_2 \right\rangle (\theta_1-\theta_2) =\left\langle \mathcal{O}_2 \mathcal{O}_2 \right\rangle (\theta_2-\theta_1) \sim 
 \frac{(l+t)^3}{4\cdot64 \pi ^2 t^3 \epsilon ^4}, \\
&\left\langle \mathcal{O}_3 \mathcal{O}_3 \right\rangle (\theta_1-\theta_2) =\left\langle \mathcal{O}_3 \mathcal{O}_3 \right\rangle (\theta_2-\theta_1) \sim 
\frac{(l+t)^3}{4\cdot64 \pi ^2 t^3 \epsilon ^4}, \\
&\left\langle \mathcal{O}_4 \mathcal{O}_4 \right\rangle (\theta_1-\theta_2)=\left\langle \mathcal{O}_4 \mathcal{O}_4 \right\rangle (\theta_2-\theta_1) \sim 
\frac{(l+t) \left(l^2-4 l t+7 t^2\right)}{4\cdot64 \pi ^2 t^3 \epsilon ^4}, \\
%%%%%%%%%%%%%%%%%%%%%%%%%%%%%%%%%%%%%%%%%%%%%%%%%%%%%%%%%%%%%
&\left\langle \mathcal{O}_1 \mathcal{O}_1 \right\rangle (\theta_1-\theta_2+2\pi)= \left\langle \mathcal{O}_1 \mathcal{O}_1 \right\rangle (\theta_2-\theta_1-2\pi) \\
&=\left\langle \mathcal{O}_1 \mathcal{O}_1 \right\rangle (\theta_1-\theta_2-2(n-1)\pi)= \left\langle \mathcal{O}_1 \mathcal{O}_1 \right\rangle (\theta_2-\theta_1+2(n-1)\pi)\sim -\frac{(l-t)^3}{4\cdot64 \pi ^2 t^3 \epsilon ^4} \\
 &\left\langle \mathcal{O}_2 \mathcal{O}_2 \right\rangle (\theta_1-\theta_2+2\pi)= \left\langle \mathcal{O}_2 \mathcal{O}_2 \right\rangle (\theta_2-\theta_1-2\pi) \\
&=\left\langle \mathcal{O}_2 \mathcal{O}_2 \right\rangle (\theta_1-\theta_2-2(n-1)\pi)= \left\langle \mathcal{O}_2 \mathcal{O}_2 \right\rangle (\theta_2-\theta_1+2(n-1)\pi)\sim  -\frac{(l-t) \left(l^2+4 l t+7 t^2\right)}{4\cdot64 \pi ^2 t^3 \epsilon ^4}\\
&\left\langle \mathcal{O}_3 \mathcal{O}_3 \right\rangle (\theta_1-\theta_2+2\pi)= \left\langle \mathcal{O}_3 \mathcal{O}_3 \right\rangle (\theta_2-\theta_1-2\pi) \\
&=\left\langle \mathcal{O}_3 \mathcal{O}_3 \right\rangle (\theta_1-\theta_2-2(n-1)\pi)= \left\langle \mathcal{O}_3 \mathcal{O}_3 \right\rangle (\theta_2-\theta_1+2(n-1)\pi)\sim  -\frac{(l-t) \left(l^2+4 l t+7 t^2\right)}{4\cdot64 \pi ^2 t^3 \epsilon ^4}\\
&\left\langle \mathcal{O}_4 \mathcal{O}_4 \right\rangle (\theta_1-\theta_2+2\pi)= \left\langle \mathcal{O}_4 \mathcal{O}_4 \right\rangle (\theta_2-\theta_1-2\pi) \\
&=\left\langle \mathcal{O}_4 \mathcal{O}_4 \right\rangle (\theta_1-\theta_2-2(n-1)\pi)= \left\langle \mathcal{O}_4 \mathcal{O}_4 \right\rangle (\theta_2-\theta_1+2(n-1)\pi)\sim -\frac{(l-t)^3}{4\cdot64 \pi ^2 t^3 \epsilon ^4} \\
\end{split}
\ee
The green functions $\left\langle\mathcal{O}_i\mathcal{O}_{j\neq i}\right \rangle$ vanish.

%%%%%%%%%%%%%%%%%%%%%%%%%%%%%%%%%%%%%%%


\begin{thebibliography}{}
\bibitem{MNF}
We prepare the paper for more precise relation between the effective reduced density matrix and diagrams.
\end{thebibliography}
\end{document}